\documentclass[traditabstract,bibyear]{aa} 
\usepackage{txfonts}
\usepackage{natbib}
\usepackage{graphicx}
\usepackage{xcolor,ulem}

\begin{document}

   \title{The high-redshift gamma-ray burst GRB\,140515A}

   \subtitle{A comprehensive X-ray and optical study}

   \author{A. Melandri\inst{1}, M. G. Bernardini\inst{1}, P. D'Avanzo\inst{1}, R. S\'{a}nchez-Ram\'{i}rez\inst{2,3,4}, F. Nappo\inst{5,1}, L. Nava\inst{6}, \\
   J. Japelj\inst{7}, A. de Ugarte Postigo\inst{2,8}, S. Oates\inst{2}, S. Campana\inst{1}, S. Covino\inst{1}, V. D'Elia\inst{9,10}, G. Ghirlanda\inst{1}, E. Gafton\inst{11}, \\ 
   G. Ghisellini\inst{1}, N. Gnedin\inst{13,14,15}, P. Goldoni\inst{12}, J. Gorosabel\inst{2,3,4}, T. Libbrecht\inst{11}, D. Malesani\inst{8}, R. Salvaterra\inst{16}, \\
   C. C. Th\"one\inst{2}, S. D. Vergani\inst{17,1}, D. Xu\inst{18,8}, G. Tagliaferri\inst{1}}


%

   \institute{
   $^{1}$ INAF - Osservatorio Astronomico Brera, Via E. Bianchi 46, I-23807, Merate (LC), Italy\\
              \email{andrea.melandri@brera.inaf.it}\\
   $^{2}$ Instituto de Astrof\'{i}sica de Andaluc\'{i}a (IAA-CSIC), Glorieta de la Astronom\'{i}a s$/$n, E-18008 Granada, Spain\\
   $^{3}$ Unidad Asociada Grupo Ciencias Planetarias (UPV/EHU, IAA-CSIC), Departamento de F\'{i}sica Aplicada I, E.T.S. Ingenier\'{i}a, Universidad del Pa\'{i}s Vasco (UPV/EHU), Alameda de Urquijo s/n, E-48013 Bilbao, Spain\\
   $^{4}$ Ikerbasque, Basque Foundation for Science, Alameda de Urquijo 36-5, E-48008 Bilbao, Spain\\
   $^{5}$ Universit\'{a} degli Studi dell'Insubria, via Valleggio 11, I-22100 Como, Italy\\
   $^{6}$ Racah Institute of Physics, The Hebrew University of Jerusalem, Jerusalem 91904, Israel\\
   $^{7}$ Faculty of Mathematics and Physics, University of Ljubljana, Jadranska ulica 19, SI-1000 Ljubljana, Slovenia\\
   $^{8}$ Dark Cosmology Centre, Niels Bohr Institute, University of Copenhagen, Juliane Maries Vej 30, DK-2100 Copenhagen, Denmark\\
   $^{9}$ ASI - Science Data Center, Via del Politecnico snc, I-00133 Roma, Italy\\
   $^{10}$ INAF - Osservatorio Astronomico di Roma, Via Frascati 33, I-00040 Monte Porzio Catone (RM), Italy\\
   $^{11}$ The Oskar Klein Center, Department of Astronomy, AlbaNova, Stockholm University, 10691 Stockholm, Sweden\\
   $^{12}$ APC, U. Paris Diderot, CNRS/IN2P3, CEA/IRFU, Obs. Paris, Sorbonne Paris Cit\'{e}, France\\
   $^{13}$ Particle Astrophysics Center, Fermi National Accelerator Laboratory, Batavia, IL 60510, USA\\
   $^{14}$ Department of Astronomy \& Astrophysics, The University of Chicago, Chicago, IL 60637 USA \\
   $^{15}$ Kavli Institute for Cosmological Physics, The University of Chicago, Chicago, IL 60637 USA \\
   $^{16}$ INAF - IASF Milano, Via E. Bassini 15, I-20133, Milano, Italy\\
   $^{17}$ GEPI, Observatoire de Paris, CNRS, Univ. Paris Diderot, 5 place Jule Janssen, F-92190 Meudon, France\\
   $^{18}$ National Astronomical Observatories, Chinese Academy of Sciences, 20A Datun Road, Chaoyang District, 100012 Beijing, China\\
             }

   \date{}

 
  \abstract{High-redshift gamma-ray bursts have several advantages for the study of the distant universe, providing unique information about the structure and properties of the galaxies in which they exploded. Spectroscopic identification with large ground-based telescopes has improved our knowledge of the class of such distant events. We present the multi-wavelength analysis of the high-$z$ {\it Swift} gamma-ray burst GRB\,140515A ($z = 6.327$). The best estimate of the neutral hydrogen fraction of the intergalactic medium (IGM) towards the burst is $x_{HI} \leq 0.002$. The spectral absorption lines detected for this event are the weakest lines ever observed in gamma-ray burst afterglows, suggesting that GRB\,140515A exploded in a very low density environment. Its circum-burst medium is characterised by an average extinction (A$_{\rm V} \sim 0.1$) that seems to be typical of $z \ge 6$ events. The observed multi-band light curves are explained either with a very flat injected spectrum ($p = 1.7$) or with a multi-component emission ($p = 2.1$). In the second case a long-lasting central engine activity is needed in order to explain the late time X-ray emission. The possible origin of GRB\,140515A from a Pop III (or from a Pop II stars with local environment enriched by Pop III) massive star is unlikely.}
  

\authorrunning{A. Melandri et al. 2015}
\titlerunning{GRB\,140515A}

   \maketitle
%

\section{Introduction}

A better understanding of the chemical enrichment and evolution of the high-redshift universe is one of the fundamental goals of modern astrophysics. High redshift surveys have been performed by means of wide field surveys of bright quasars \citep[e.g.][]{fan12} or deep field analyses to identify distant galaxies by their drop-out \citep[e.g.][]{bou14}. The identification of high-redshift Gamma-Ray Bursts (GRBs) add a different and profitable view of the distant universe \citep[see][for a recent review]{sal15}. With respect to other probes, GRBs have many advantages: (i) they are detected at higher redshifts; (ii) they are independent on the galaxy brightness; (iii) they do not suffer of usual biases affecting optical/NIR surveys; (iv) they reside in average cosmic regions. High-$z$ GRBs can provide fundamental, and in some cases unique, information about the early stages of structure formation and the properties of the galaxies in which they blow up. For example, GRBs can be used to trace the cosmic star formation rate (Kistler et al. 2009; Ishida et al. 2011; Roberson \& Ellis 2012), to pinpoint high-$z$ galaxies and explore their metal and dust content (Tanvir et al. 2012; Salvaterra et al. 2013; Elliott et al. 2015), and to shed light on  the re-ionization history (Gallerani et al. 2008; McQuinn et al. 2008), to constrain the dark matter particle mass (de Souza et al. 2013) and the amount of non-Gaussianity present in the primordial density field (Maio et al. 2013), and to measure the level of the local inter-galactic radiation field (Inoue et al. 2010). Additionally, they could also provide direct and/or indirect evidences for the existence of the first, massive, metal-free stars, the so-called Population~III stars (Campisi et al. 2011; Toma et al. 2011; Wang et al. 2012; Ma et al. 2015). 

Since the launch of the \textit{Swift} satellite \citep{ge04} 8 events have been identified at redshift greater than $\sim 6$, and for 5 of them spectroscopic redshift was secured, including in the list \object{GRB\,140515A} that we are discussing in this paper. Remarkably, some of them showed fairly bright early-time afterglows, even detectable by small robotic telescopes \citep[e.g. GRB\,050904; ][]{Tagl05,boer}, but in general the observational features of high-$z$ events do not seem to differ significantly from those of their closer siblings, as clearly pointed out studying, e.g., the event with the highest spectroscopically confirmed redshift, $z\sim8.2$: GRB\,090423 \citep{sal09,Tan09}.

In this paper we describe our observations of high-$z$ GRB\,140515A in Sect.\,\ref{sec:obs} and the result of the analyses in Sect.\,\ref{sec:res}. A discussion is reported in Sect.\,\ref{sec:dis} and the main conclusions of our work are summarised in Sect.\,\ref{sec:concl}.

Throughout the paper, distances are computed assuming a $\Lambda$CDM-universe with H$_0 = 71$\,km\,s$^{-1}$\,Mpc$^{-1}$, $\Omega_{\rm m}$ = 0.27, and $\Omega_\Lambda$ = 0.73 \citep{larson,Kom11}. Magnitudes are in the AB system and errors are at $1\sigma$ confidence level. Raw and reduced data not explicitly reported in tables are available from the authors upon request.

\section{Observations}
\label{sec:obs}

On 2014 May 15 at 09:12:36 UT (= T$_{0}$), the {\it Swift}/BAT triggered and located the long GRB\,140515A \cite{paolo}. {\it Swift}/XRT promptly detected the afterglow emission whereas {\it Swift}/UVOT did not identify any credible bright optical candidate. A faint optical afterglow was later identified by ground-based observations with 8m Gemini-North telescope \cite{fong}, the 2.5m NOT telescope \cite{ugarte}, the 2.2m GROND telescope \cite{graham}, and the 3.6m TNG telescope \cite{mela}.

Spectroscopic observations performed with the Gemini-North telescope \cite{chor} and the GTC telescope \cite{ugarte2} detected sharp decrement in flux below 8900 \AA, caused by Ly$\alpha$ absorption at redshift $z$=6.327 \cite{chor2}. Spectral analysis will be described in detail in Section 3.3.1.

\section{Results}
\label{sec:res}

\subsection{BAT temporal and spectral analysis}

\begin{figure}
\centering
\includegraphics[height=8cm,width=\hsize,clip]{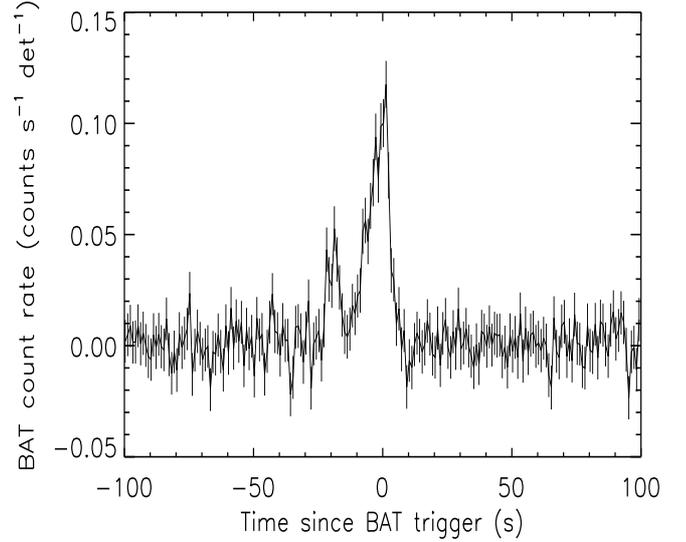}
\caption{BAT mask--weighted light curve showing the count rate in the $15-150$ keV energy range.}
\label{batlc}
\end{figure}

The \textit{Swift}/BAT data were processed with the standard \textit{Swift} analysis software included in the NASA's HEASARC software (HEASOFT, ver. 6.16) and the relevant latest calibration files. For each GRB, we extracted mask--weighted, background-subtracted light curves and spectra with the \texttt{batmaskwtevt} and \texttt{batbinevt} tasks in FTOOLS. The mask weighted light curve shows a double-peaked structure. The first pulse started at $T_{0}-22$ s and peaked at $T_{0}-18$ s. It was followed by a second brighter pulse between $T_{0}-10$ s and $T_{0}+4$ s (see Fig. \ref{batlc}). The total duration of the burst event in the $15-150$ keV energy band in the observer frame is $T_{\rm 90} = (23.4 \pm 2.1)$~s (90$\%$ confidence level), corresponding to $\sim 3.2$~s in the rest-frame.

A fit to a simple power law of the time-averaged spectrum from $T_{0}-22$ s to $T_{0}+4$ s gives a photon index $\Gamma=1.86 \pm 0.14$ ($\chi^2=61.21$, d.o.f.$=56$). A power law with an exponential cutoff gives a moderately better fit ($\chi^2=54.57$, d.o.f.$=55$; F--test probability P$=98.8\%$). For this model the photon index is $\Gamma=0.99^{+0.63}_{-0.80}$, $E_{\rm pk} = 51.8_{-22.0}^{+93.0}$ keV and the total fluence in the $15-150$ keV band is $F_{\rm BAT}=(6.53_{-0.57}^{+0.47}) \times 10^{-7}$ erg/cm$^2$. The 1-sec peak flux measured from $T_{0}+1.50$ s in the $15-150$ keV band is $f_{\rm pk,BAT}=0.86 \pm 0.10$ ph cm$^{-2}$ s$^{-1}$).

We also tested the presence of a blackbody component in the prompt emission spectrum. We added a blackbody component to the non thermal (power-law) spectrum and we fit this model to the data, obtaining a photon index $\Gamma=1.94_{-0.62}^{+1.17}$ and a blackbody temperature $kT=12.4_{-4.0}^{+5.3}$ keV, that in the source rest frame corresponds to $kT_{\rm rf}=90.3_{-29.0}^{+39.0}$ keV. This model provides an adequate fit ($\chi^2=54.29$, d.o.f.$=54$) but not a significant improvement compared to the cutoff power-law model. 

We searched for spectral evolution between the two main emission episodes of the prompt emission. The spectrum of the first peak (from $T_{0}-22$ s to $T_{0}-14$ s) can be modelled as a power-law spectrum, with a photon index $\Gamma=2.01_{-0.32}^{+0.36}$. The spectrum's second peak (from $T_{0}-14$ s to $T_{0}+4$ s) is better represented by a power law with an exponential cutoff (F--test probability P$=99.6\%$), with photon index $\Gamma=0.82^{+0.63}_{-0.75}$, $E_{\rm pk} = 52.7_{-23.4}^{+92.0}$ keV.

The total bolometric (rest frame $1-10^4$ keV) isotropic energy, assuming the exponential cutoff model, is $E_{\rm iso} = (5.8 \pm 0.6) \times 10^{52}$~erg at $z=6.327$ and $E_{\rm pk,rf}=379.7_{-161.3}^{+681.7}$ keV, consistent with the $E_{\rm pk,rf}-E_{\rm iso}$ correlation within its $1\,\sigma$ scatter \citep{EpEiso, amati, nava}. The isotropic luminosity is $L_{\rm iso} = (3.6 \pm 0.8) \times 10^{52}$~erg~s$^{-1}$, consistent with the $E_{\rm pk,rf}-L_{\rm iso}$ correlation within its $1\sigma$ scatter \cite{yone, nava}.

\subsection{XRT temporal and spectral analysis}

  \begin{figure}
   \centering
  \includegraphics[width=7.0cm,height=9.5cm,angle=270]{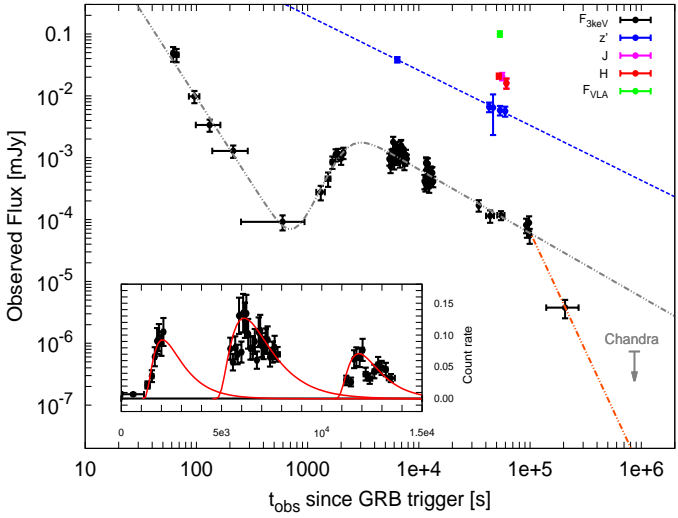}
   \caption{Observed multi-wavelengths light curve of GRB\,140515A. We report the best fits for the optical (blue) and X-rays (grey) band. The orange line shows the possible decay of the late time X-rays afterglow assuming a possible jet-break. {\it Inset}: The time interval of the X-ray bump in linear scale. Red curves represent three independent flaring episodes.}
              \label{FigLC1}
    \end{figure}

The XRT began observing $\sim$60~s after the BAT trigger, with the first 9~s in Windowed Timing (WT) mode and the remainder in Photon Counting (PC) mode. We collected the XRT data from the online Burst Analyzer \cite{evans} and converted the observed [0.3-10] keV count rate into flux at 3 keV. The X-ray light-curve at that energy (Fig. \ref{FigLC1}) is well described by an initial steep decay ($\alpha_{\rm 1} \sim 2.9$) followed by a broad bump after $\sim 10^{3}$~s that lasted for almost 1 day. From Fig. \ref{batlc} it appears clear that the ``real'' burst began at $T_{\rm 0, GRB} = T_{\rm 0, real} \sim T_{\rm 0} - 22$~s. If we consider $T_{\rm 0, GRB}$ as the beginning of the burst, shifting backwards the temporal axis of Fig. \ref{FigLC1}, the initial decay becomes less steep ($\alpha_{\rm 1} \sim 2.4$), but it makes no difference for late times breaks and decay indices (see Tab.\ref{tab1}).

At later times a further steepening in the light-curve is detected, also confirmed by Chandra deep upper limit. If we consider only XRT data, the late time decay index is $\alpha_{\rm late} = 3.9 \pm 0.6$, while if we take also into account the deep late time upper limit this value appears to be at least $\geq 2.6$ \cite{raffa}. 
Although it cannot be confirmed by the optical/NIR data, the late time X-ray decay index might suggest a possible jet-break origin. 

We performed a time-integrated spectral analysis of the X-ray emission from $t-T_{0}=86.4$ s to $t-T_{0}=10^5$ s using XSPEC ver. 12.8.2. The best fit to the data is an absorbed power law model: the Galactic absorption is kept fixed to the value $N_{H}^{\rm Gal} = 2.54\times 10^{20}$ cm$^{-2}$ \cite{willi}, the photon index is $\Gamma = (1.79 \pm 0.11)$ and the local absorption at $z=6.327$ is $N_{H} < 5.6 \times 10^{22}$ cm$^{-2}$ (c-stat=341.1, d.o.f.$=360$). Selecting photons from $t-T_{0}=86.4$ s to $t-T_{0}=2152$ s, when the spectral changes are minor, we obtain an upper limit on the column density of $N_{H} < 8.9 \times 10^{22}$ cm$^{-2}$.

Also for the XRT spectrum we tested the presence of a possible blackbody component. The resulting fit adding a thermal component to the power-law spectrum did not improve the absorbed power law model described above.

The total isotropic energy in the $0.3-30$~keV rest frame energy band is $E_{\rm X,iso}=(7.19 \pm 0.43)\times 10^{51}$~erg, thus GRB\,140515A is also consistent with the $E_{\rm X,iso}-E_{\rm pk,rf}-E_{\rm iso}$ correlation within its $2\,\sigma$ scatter \cite{3par1,3par2}.

\begin{table}
\caption{Observed X-ray light-curve fitting results ($\chi^{2}$/d.o.f. = 59.31/51 = 1.16 assuming that $T_{\rm 0, GRB} = T_{\rm 0}$, and $\chi^{2}$/d.o.f. = 55.89/51 = 1.09 if instead $T_{\rm 0, GRB} = T_{\rm 0, real}$).} 
\label{tab1}     
\centering        
\begin{tabular}{ccc}
\hline\hline             
Parameter & $T_{\rm 0, GRB} = T_{\rm 0}$ & $T_{\rm 0, GRB} = T_{\rm 0, real}$\\
\hline
$\alpha_{\rm 1}$ & 2.80 $\pm$ 0.22 & 2.34 $\pm$ 0.15 \\
T$_{\rm break}$ & 732 $\pm$ 25 ~s &729 $\pm$ 18 ~s \\
$\alpha_{\rm 2}$ & -3.87 $\pm$ 0.74 &  -4.03 $\pm$ 0.74\\
T$_{\rm peak}$ & 3028 $\pm$ 192 ~s & 2930 $\pm$ 133 ~s\\
$\alpha_{\rm 3}$ & 1.03 $\pm$ 0.05 & 1.03 $\pm$ 0.04\\
T$_{\rm jet-break}$ & $\geq 10^{5}$ ~s &$\geq 10^{5}$ ~s \\
$\alpha_{\rm late}$ & 3.9 $\pm$ 0.6 & 3.9 $\pm$ 0.6\\                      
\hline
\end{tabular}
\end{table}

\subsection{Optical/NIR temporal analysis}

 The {\it Swift}/UVOT began observing the field of GRB 140515A 3.7~ks after the trigger \cite{paolo}. The afterglow was not detected in any of 7 UVOT filters. This is consistent with the redshift reported by Chornock et al. (2014a). In order to provide deep upper limits, we co-added the exposures within the first sequence of observations (00599037000). We determined the count rate using a 5 arcsec circular source region centred at the optical afterglow position reported by Fong et al. (2014), and a circular background region of radius 20 arcsec positioned on a blank area of sky situated near to the source position. The photometry was extracted using the UVOT tool {\it uvotsource}. The count rates were converted to magnitudes using the UVOT photometric zero points \cite{bree}. We used Heasoft software version 6.15.1 and UVOT calibration version 20130118. 
 
 In Tab. \ref{tab2} we summarise ultraviolet and optical 3$\sigma$ upper limits and other optical/near infrared detections of the optical afterglow. We note that at the redshift of GRB\,140515A the SDSS-$z$ filter is slightly affected by the absorption of the intergalactic medium (IGM) since the optical depth for the Ly-$\alpha$ at $z > 6$ rises dramatically (i.e. Fan et al. 2006). Extrapolating the IGM properties from low redshift to $z = 6.327$ we estimated the expected correction for the $z'$ filter to be $\ge$ 0.30$^{+0.08}_{-0.03}$~mag (see Japelj et al. 2012 for details of the method). Despite the fact that this is formally a lower limit of the correction - not taking into account the rise in optical depth at $z > 6$ - we took it into account before converting all the observed magnitudes reported in Tab. \ref{tab2} into flux densities.
 
Although the optical light curve is sparsely sampled (Fig. \ref{FigLC1}) it is possible to estimate the decay index of the optical afterglow in the $z'$-band at late times. From 6~ks after the burst event the optical afterglow follows a power-law decay with $\alpha_{\it z'} = 0.89 \pm 0.02$.

 \begin{table}
\caption{Optical observations. Magnitudes are in AB system and have not been corrected for Galactic absorption along the line of sight \citep[E$_{\rm (B-V)} = 0.02$~mag, ][]{schla}. References for data taken from the GCNs are: 1) Fong et al. 2014; 2) Graham et al. 2014. }
\label{tab2}     
\centering        
\begin{tabular}{ccccc}
\hline\hline             
T$_{\rm mid}$ & Exposure & Filter & Mag & Ref.\\
\hline
[sec] & [sec] & & &\\
\hline
\hline
3743.0  &   841    &  $white$    &  $>$ 22.14 & UVOT \\
6795.0  &    393   &  $uvw2$   &    $>$ 20.73 & UVOT \\
9203.0  &   1082  &   $uvm2$   &    $>$ 20.90 & UVOT \\
6885.0  &   1141  &   $uvw1$   &   $>$ 21.28 & UVOT \\
6982.0  &   1437  &   $u$   &  $>$ 21.35 & UVOT \\
3875.0  &   549    &  $b$    &   $>$ 21.13 & UVOT \\
4220.0  &   568    &   $v$   &   $>$ 20.27 & UVOT  \\
6408.0 & 480 & $z'$ & 20.27 $\pm$ 0.11 & 1 \\
42877.1 & 1500 & $z'$ & 22.18 $\pm$ 0.19 & NOT \\
46462.8 & 60 & $z'$ & 22.21 $\pm$ 0.35 & X-shooter \\
53650.4 & 1500 & $z'$ & 22.32 $\pm$ 0.19 & NOT \\
59705.5 & 1500 & $z'$ & 22.35 $\pm$ 0.20 & NOT \\
61200.0 & 3000 & $z'$ & 22.1 $\pm$ 0.1 & 2 \\
56160.0 & 1800 & $J$ & 20.63 $\pm$ 0.15 & TNG \\
61200.0 & 2400 & $J$ & 20.9 $\pm$ 0.2 & 2 \\
52344.0 & 3600 & $H$ & 20.61 $\pm$ 0.10 & TNG \\
61200.0 & 2400 & $H$ & 20.9 $\pm$ 0.2 & 2 \\
\hline
\end{tabular}
\end{table}

 \subsection{Optical/NIR spectral analysis}
 
 \subsubsection{GTC spectrum}
 
We obtained spectroscopy of the afterglow of GRB\,140515A with OSIRIS \cite{cepa} at the 10.4m Gran Telescopio Canarias \cite{ugarte2}. The observations were obtained between 22:37:31 UT and 00:09:46 UT (mean epoch 14.184 hr after the GRB onset) with 0.6$^{\prime\prime}$ seeing and consisted of $3\times1800$ s exposures. We used the R2500I VPH grism, which covers the range between 7330 and 10000 \AA~at a resolution of $\sim$1600 using a 1$^{\prime\prime}$ slit. 

The data were reduced in a standard way (bias subtraction, pixel-to-pixel response correction, cosmic ray removal, wavelength calibration, 1D extraction, flux calibration, and combination of spectra) using self-made routines based on IRAF \cite{tody}. The resulting combined GTC spectrum shows a strong continuum above $\sim$ 8900 {\AA}, where the signal-to-noise ratio is $\sim20$ per pixel, or $\sim40$ per resolution element. 

 \subsubsection{X-shooter spectrum}
  
 We observed the field of GRB\,140515A with the X-shooter spectrograph mounted at the ESO/VLT using the nodding mode with 1 $\times$ 2 binning. The spectrum was acquired on 2014 May 16, starting at 00:42:43 UT ($\sim$ 15.5~hr after the GRB onset) and consisted of 2x4x600~s exposures, for a total integration time of 4800~s on source, covering the range between $\sim$3000 and $\sim$ 24000 \AA. The mid expose time is 16.3~hr ($\sim$ 0.68~d) after the GRB trigger. The final reduced spectrum has a signal-to-noise ratio of $\sim$ 3 per pixel\footnote{The quoted difference in S/N  between the GTC and Xshooter spectra is due partly to the different pixel size of the two instruments and partly to the better observing conditions of the GTC observation.}, with a seeing of $\sim$ 0.9$^{\prime\prime}$ (measured from combined 2D spectrum in the VIS and NIR arms). The flux calibration of the X-shooter, which is problematic in general (Kr\"{u}hler et al. 2015; Japelj et al. 2015), is uncertain due to unavailable standard spectrophotometric star in the night when the observations were done and because the photometric observations, which could be used to check the quality of calibration, have rather high errors at this epoch. Thus, as the photometric observations at this epoch have rather high errors, it is not possible to use them to reliably rescale the spectrum.

\subsubsection{Ly$\alpha$ forest constraints on the IGM}

\begin{figure}
\hspace{-0.6cm}
\includegraphics[width=0.40\textwidth,angle=270]{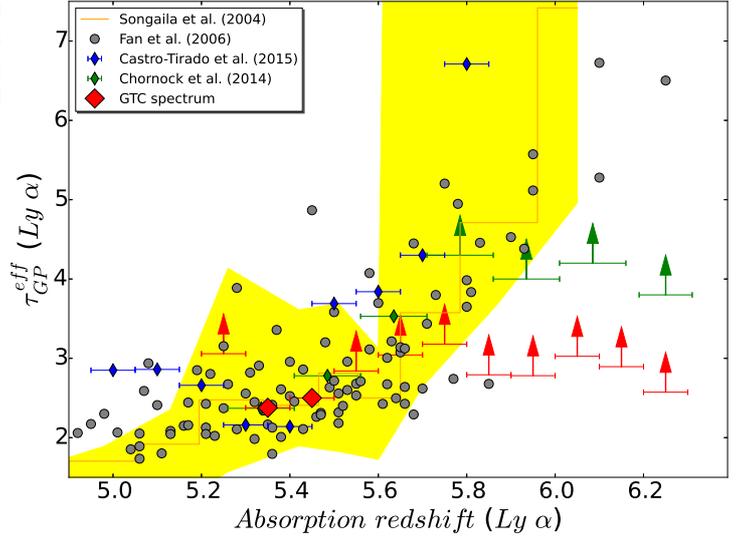}
\caption{Ly$\alpha$ effective optical depth in the line of sight of GRB\,140515A compared with previous GRB and QSO works. The coloured area shows the optical depth found by Songaila (2004) while grey points are measurements from Fan et al. (2006) with sample of quasars.}
\label{fig:od}
\end{figure}

We analysed the ionisation state of the IGM using the Gunn $\&$ Peterson (1965) optical depth, defined as $\tau^{eff}_{GP}=-\ln(\mathcal{T})$, where $\mathcal{T}$ the average transmission in a redshift bin. Following Songaila $\&$ Cowie (2002) and Songaila (2004) we normalised the GTC spectrum (as its signal-to-noise ratio, SNR, is better than the X-shooter one, see section 3.4.4) by fitting a power law to the continuum, and divided it into redshift bins of 0.1 between $z=5.2$ and $z=6.3$. The results are presented in Fig. \ref{fig:od} and in Table \ref{tab:od}. 

We only see sky line residuals up to $z\sim5.5$, above which we can just give detection limits based on the noise spectrum. Our limits are less restrictive than the ones presented by Chornock et al. (2014b) due to the lower SNR, but show the same behaviour (Fig. \ref{fig:od}). Results coming from both GRB\,140515A and GRB\,130606A \citep{chor3,castro,olga} are consistent with quasar measurements \cite{song2,fan}.

\subsubsection{Ly$\alpha$ red damping wing fitting}

\begin{figure}
\hspace{-0.4cm}
\includegraphics[width=0.41\textwidth,angle=270]{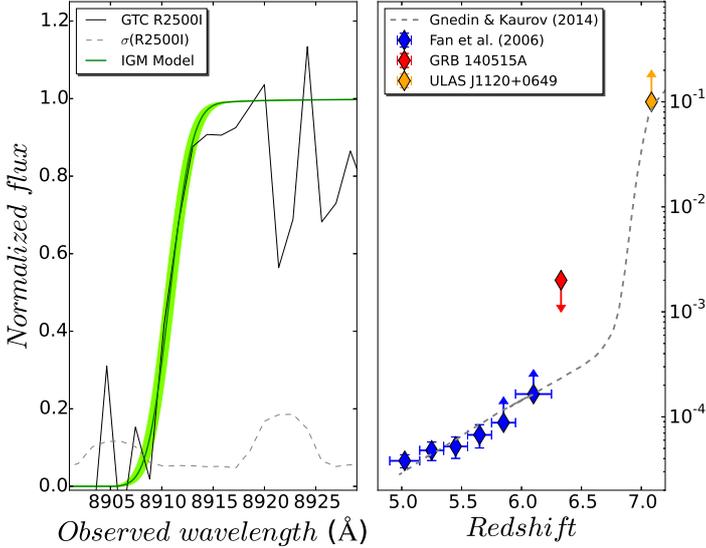}
\caption{\textit{Left}: Best IGM damping wing fit to the spectrum of GRB 140515A. \textit{Right}: Redshift evolution of the hydrogen neutral fraction. The dotted line shows the Gnedin $\&$ Kaurov (2014) model and points (see legend) the observational measurements of this quantity. Points with arrows are lower/upper limits.}
\label{fig:xhi}
\end{figure}

We tried to fit the strongest feature seen in the spectrum (at $\sim 8900~\AA$) to an absorption Lyman-$\alpha$ feature with a Voigt profile. Following Chornock et al. (2014b), we first computed a Voigt model using the same constraints, obtaining inconsistent results. This could be due to the fact that they do not seem to consider the instrumental profile, whose effect on the Ly-$\alpha$ feature is not negligible at this resolution when $\log(N_{\rm HI}) \lesssim 19$. Looking at Fig. \ref{fig:xhi}, we can observe the residuals of a sky line subtraction few angstroms blue-wards the wing, precisely at the zone crucial to fit a Voigt model. After a careful inspection on the 2D images of both GTC and X-shooter instruments, we concluded that there is no flux at this zone. Consequently, the wing profile is too sharp to get a satisfactory fit, suggesting that the absorption is dominated by the IGM and that the host absorption is masked.

We then built up IGM models following the prescription of Miralda-Escud{\'e} (1998), fixing the lower redshift value to $z=6.0$ because the contribution to the wing shape below this redshift is negligible (it starts to be important closer to the host). Our best fit, with $z=6.3298 \pm 0.0004$~~and a fraction of neutral hydrogen $x_{HI} \leq 0.002$, is shown in Fig. \ref{fig:xhi}. We caution that due to the sharpness of the wing, the few points we have because of GTC resolution, and the sky line next to the absorption, any formal constraints on these quantities would be unreliable, so the values should be interpreted as the most plausible estimations that we can obtain from the data. Moreover, especially by the fact that $z$ cannot be determined by metal lines, hybrid models cannot offer a more accurate fit than the one showed in Fig. \ref{fig:xhi}, so no constraints on the host HI abundance can be derived from this event \citep[for further discussion, see][]{miralda}. However, due to the sharpness of the red damping wing, it is obvious that the neutral hydrogen present in the IGM cannot mask neither the presence of a DLA nor a subDLA, as their damping wings would be easily identified. Consequently, we can establish a conservative upper limit of $\log(N_{\rm HI}) \lesssim 18.5$ for the HI abundance in the host galaxy of GRB\,140515A. As shown in Fig. \ref{fig:xhi}, the fraction of neutral hydrogen derived from this analysis is in good agreement with the model by Gnedin $\&$ Kaurov (2014), and it provides a very relevant observational constraint. 

Last, we estimated the 3$\sigma$ upper limits on the observer-frame equivalent width (EW) for the Si II $\lambda$1260, O I  $\lambda$1302, and C II $\lambda$1334. We find a value of 0.67 \AA, 1.06 \AA, and 1.30 \AA, respectively. These estimates are a factor $\sim2$ more stringent of what reported by Chornock et al. (2014b), resulting to upper limits on the gas-phase abundances of [Si/H] $\lesssim -1.4$, [O/H] $\lesssim -1.1$, and [C/H] $\lesssim -1.0$. Furthermore, these lines are weaker than the average rest-frame EWs observed for a typical GRB \cite{ugarte0}. In fact, the strength of those lines compared to the average GRB spectrum that can be estimated with the use of the line strength parameter (LSP, as defined in de Ugarte Postigo et al. 2012), is LSP~$<-3.15$, $<-3.89$, and $<-2.88$, respectively. This means that these lines are very weak and that GRB\,140515A exploded in a relatively low density environment. However, our limits on the metals abundances do not allow us to put a stringent limit on the metallicity of the progenitor. 


 \begin{table}
\caption{IGM absorption towards GRB 140515A.}
\centering
\begin{tabular}{ccccc}
 \hline
 \hline
 $z$ & $\mathcal{T}$ & lim($\mathcal{T}$) & $\tau^{eff}_{GP}$ & lim($\tau^{eff}_{GP}$) \\
 \hline
 5.25 &  -- &  0.0594 &  -- &  2.82 \\
 5.35 &  0.1174 &  0.0709 &  2.14 &  2.65 \\
 5.45 &  0.1038 &  0.0767 &  2.27 &  2.57 \\
 5.55 & -- &  0.0739 & -- &  2.61 \\
 5.65 & -- &  0.0604 & -- &  2.81 \\
 5.75 &  -- & 0.0527 & -- &  2.94 \\
 5.85 &  -- &  0.0775 & -- &  2.56 \\
 5.95 &  -- &  0.0784 & -- &  2.55 \\
 6.05 &  -- &  0.0614 & -- &  2.79 \\
 6.15 &  -- &  0.0700 & -- &  2.66 \\
 6.25 &  -- &  0.0965 & -- &  2.34 \\
 \hline
\label{tab:od}
\end{tabular}
\end{table}

 \subsubsection{Spectral energy distribution}
 
\begin{figure}
\centering
\includegraphics[height=7cm,width=\hsize,clip]{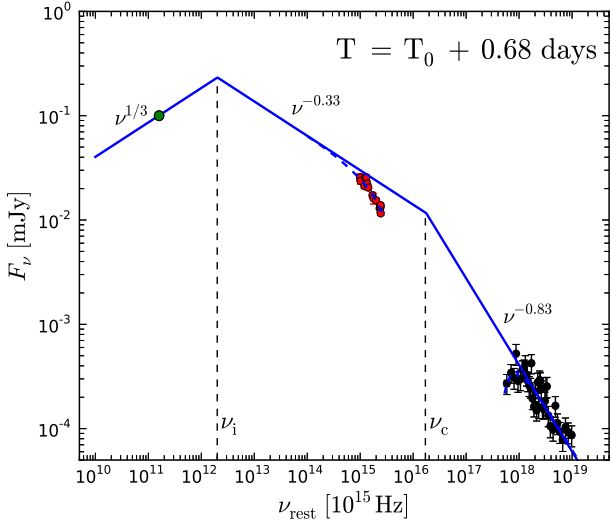}
\caption{Spectral energy distribution obtained with X-shooter (red) and XRT data (black). We included also the radio detection (green) of Laskar et al. (2014). Dashed lines mark the position of the injection frequency ($\nu_{i} = 2 \times 10^{12}$~Hz) and the cooling frequency ($\nu_{c} \sim 2 \times 10^{16}$~Hz) expected for a pure synchrotron model.}
\label{xsSED}
\end{figure}

 We constructed a broadband spectral energy distribution (SED) using the flux-calibrated X-shooter spectrum and {\it Swift} X-ray data (Fig. \ref{xsSED}). X-shooter data are treated as outlined in Japelj et al. (2015). The spectrum was corrected for Galactic extinction (E$_{\rm (B-V)}$ = 0.02 mag) by using the Cardelli etl al. (1989) extinction curve and Galactic extinction maps \cite{schla}. Regions of telluric absorption were masked out. Spectrum was re-binned in bins of approximately 50 $\mathrm{\AA}$ to increase S/N and to guarantee a comparable weight of the optical and X-ray SED part. The absolute flux calibration was fine-tuned with simultaneously obtained near-infrared photometric observations. The X-ray part of the SED was built from time integrated observations obtained between 5-60 ks and its mean epoch was interpolated to the mean epoch of the X-shooter observations.

The SED fitting was carried out with the spectral fitting package XSPECv12.8 \cite{arnaud}. We model the SED with either a single or broken power-law intrinsic spectrum - for the latter we assume $\beta_{\rm X} = \beta_{\rm O} + 0.5$ (Sari, Piran $\&$ Narayan 1998). Extinction is modelled with the three commonly assumed extinction curves of Milky Way, Large and Small Magellanic Cloud \cite{pei}. Only the spectrum red-ward of Ly$\alpha$ line has been used in analysis. The broadband SED is best described by a broken power-law and an SMC-type extinction with $A_{\rm V} = 0.11 \pm 0.02$~mag and $\beta_{\rm O} = 0.33 \pm 0.02$. Requiring $\Delta \beta = 0.5$ it has been possible to better constrain the X-ray column density to $N_{H} = 1.35_{-1.08}^{+1.22} \times 10^{22}$ cm$^{-2}$. This value is consistent with the direct estimate from X-ray data only (see Section 3.2). 

As a consistency check we also built the spectral energy distribution of the optical afterglow using all the available photometric observations reported in Table \ref{tab2} and {\it Swift} X-ray data. The extinction estimated with photometric information (A$_{\rm V} \sim 0.6$~mag) is much higher than the one obtained from the more accurate spectral analysis reported above. This is due to the fact that photometric measurements are too sparse to give reliable results. The rest-frame extinction (A$_{\rm V} \sim 0.1$~mag) of GRB\,140515A is consistent with the A$_{\rm V}$ distribution found for the complete BAT6 sample \cite{covino} and suggests that this is a typical value also for high-$z$ events.

\section{Discussion}
\label{sec:dis}

\subsection{Late time flare emission / refreshed shock}

\begin{figure}
\centering
\hspace{-0.8cm}
\includegraphics[width=\hsize,height=7.0cm]{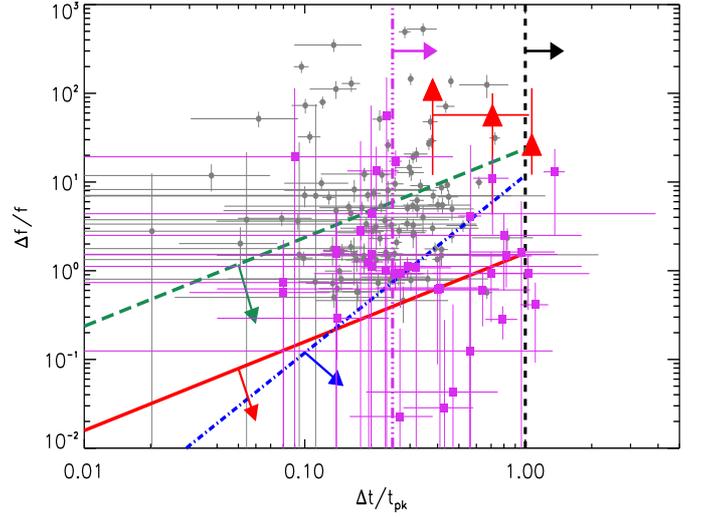}
\caption{Kinematically allowed regions for afterglow variability in the $\Delta f/f$ vs. $\Delta t/t_{\rm pk}$ plane. Coloured lines with arrows represent the allowed regions for density fluctuations on-axis (blue), density fluctuations off-axis (red), multiple density fluctuations off-axis (green), refreshed shocks (pink) and patchy shell (black), respectively \citep[see][for details]{ioka, chinca10, mgb}. In this plot we show early time ($t_{pk}\lesssim 1$ ks, grey points) and late time ($t_{pk}\gtrsim 1$ ks, magenta squares) flares. The red triangles are the three flaring episodes in GRB 140515A. The error bars account for the uncertainty on the behaviour of the underlying continuum: the lower bar corresponds to a flat power-law decay after $10^{3}$~s, the high bar to a flat decay normalised to the last datapoint, and the central value to a power-law decay consistent with the slope of the optical light curve after $10^{3}$~s.}
\label{Ioka}
\end{figure}

The most straightforward explanation for the observed X-ray peak at $t\sim3000\,$s (Fig.  \ref{FigLC1}) is the onset of the afterglow emission. This interpretation is supported by the rise and decay indices of the X-ray light curve, that are consistent with the expectations for the emission of the forward shock interacting with an homogeneous medium. An alternative possibility is that the X-ray peak corresponds to a late time flaring activity, or to variability of the GRB afterglow interacting with the ambient medium. A powerful tool to investigate the nature of this variability is the comparison of the flux increase as a function of the temporal variability of the peak with the regions of allowance for bumps in the afterglow on the basis of kinematic arguments \cite{ioka}. In Fig.~\ref{Ioka} we portrayed the sample of early time ($t_{pk}\lesssim 1$ ks) flares \cite{chinca10} and late time ($t_{pk}\gtrsim 1$ ks) flares \cite{mgb}. 

If we interpret the broad X-ray bump of GRB\,140515A as a single long-lasting flaring episode it would occupy a different region with respect to the observed X-ray flares since it is characterised by a very long duration ($\Delta t/t_{pk} \sim 50 \gg 1$) and large flux variation ($\Delta f/f \sim 10^{2}$). It would be therefore consistent with being produced by refreshed shocks \cite{rm98,kp00b,sm00}, or by an intrinsic angular structure on the emitting surface (a ``patchy shell'') \cite{mrw98,kp00a}, or by shock reflection generated by the interaction of the reverse shock with dense shells formed at
an earlier stage of the explosion \cite{hascoet}. An increase of the external medium density would require a sharp and large jump in a uniform density profile to produce the observed increase in the observed light curves, which seems unlikely. 

A single X-ray broad peak is well outside the region of validity for the internal shock model. However, there is still the possibility that the broad single peak that we are observing is the result of the superposition of multiple peaks, each with $\Delta t/t_{pk}\ll 1$. In the inset of Fig. \ref{FigLC1} we sketched a possible temporal behaviour for GRB\,140515A, where three flaring events \citep[with a typical profile as described in][]{norris}, superposed to the underlying temporal decay, could be responsible of the shape of the broad bump observed. If we consider this scenario the observed behaviour becomes consistent with the internal shocks scenario (Fig. \ref{Ioka}). This situation resembles the case of GRB\,050904, a GRB at very similar redshift ($z=6.29$) that shows a late time variability in the X-rays and a sudden drop of the observed emission afterwards. However, GRB 140515A is fainter than GRB\,050904 and its variability has not been fully captured by the XRT.



\subsection{Standard afterglow interpretation}

 \begin{figure}
\centering
\hspace{-0.7cm}
\includegraphics[height=7cm,width=9.5cm,clip]{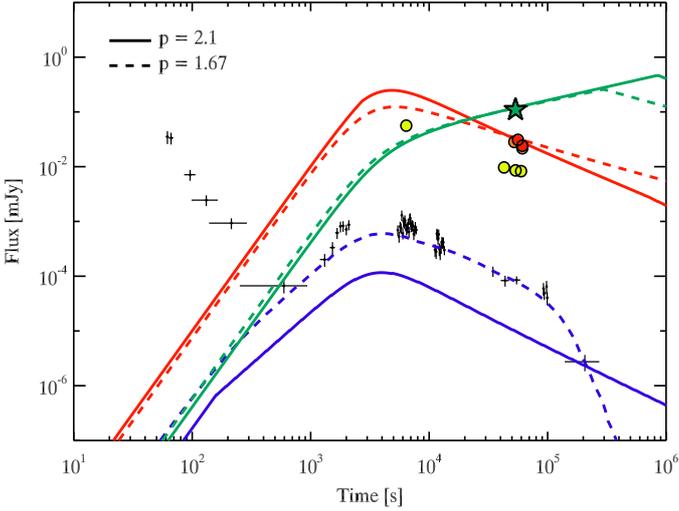}
\caption{Two multi--wavelength interpretations of GRB 140515A afterglow light curve in X-ray band (black dots, blue lines), SDSS--$z'$ band (yellow circles), $J$ band (orange circles), $H$ band (red circles, red lines) and radio band (green star, green lines). Dashed lines represent the solution in the ``flat electron spectrum''  scenario ($p = 1.67$). Solid lines describe the solution for the ``multi--component'' scenario ($p = 2.1$).}
\label{afterglow}
\end{figure}

In order to explain the observed light curves, we consider a semi--analytic model that describes the dynamical evolution of the fireball when interacting with the external circumburst medium and the respective radiative emission in the standard forward shock scenario \cite{nava2}. The radiative description is based on the model illustrated by Nappo et al. (2014), that allows us to compute the synchrotron spectrum as a function of time, normalised to the bolometric luminosity obtained by the dynamical model. We assume that the electrons are injected with a power--law energetic distribution with index $p$ and can cool for synchrotron and synchrotron self-Comtpon (SSC) radiation. The model allows to obtain at each step:

\begin{itemize}
\item[(i)] the synchrotron break frequencies (i.e. the self--absorption frequency $\nu_{\rm a}$, 
the injection frequency $\nu_{\rm i}$ and the cooling frequency $\nu_{\rm c}$)
\item[(ii)] the fraction of dissipated energy that is emitted in radiation $\epsilon_{\rm rad}$ 
that is used to determine the bolometric luminosity.
\item[(iii)] the comptonization parameter $Y$
\item[(iv)] the synchrotron spectrum $L_{\rm \nu, syn}$
\end{itemize}

Since the spectrum of the radiation is estimated at each time, the light curve at a specific frequency $F_{\nu}(t_{\rm obs})$ can be derived.

The parameters that can be varied in order to reproduce the observed light curves are: the initial bulk Lorentz factor ($\Gamma_0$), the isotropic prompt emitted energy $E_{\rm iso}$, the prompt radiation efficiency ($\eta$), the circumburst density ($n$, in case of homogeneous medium), the injected electron spectral index ($p$), the fraction of dissipated energy distributed to the leptons ($\epsilon_{\rm e}$) and to the magnetic field ($\epsilon_{\rm B}$). The model assumes an isotropic ejecta, so it can not reproduce geometrical features as jet break and side-expansion effect.

\begin{table} 
\centering
\caption{ 
Parameters of the proposed light curves scenarios. We estimated $\theta_{\rm jet}$ from Equation 1 in Sari, Piran $\&$ Halpern (1999) and Ghirlanda, Ghisellini $\&$ Lazzati (2004), using $E_{\rm iso}=6\times10^{52}\,$erg.}
\begin{tabular}{cccc}
  \hline \hline
  & & flat spectrum & multi--component \\
  \hline
$\Gamma_0$ & & $195$       	&$125$	\\
$\eta$   &     &$0.02$	&$0.04$	\\
$n$  & [${\rm cm}^{-3} $]     & $0.05$ & $0.5$ \\
$\epsilon_{\rm e}$ & & $0.15$	&$0.12$\\
$\epsilon_{\rm B}$ & &  $2.3 \times 10^{-4}$	&$1.5 \times 10^{-4}$\\
$p$ & & $1.67$	&$2.1$\\
$\theta_{\rm jet}$ & [$^{\circ}$] & $>$1.5 & $>$2.9\\
$E_{\gamma}$ & [erg] & $>2.1 \times 10^{49}$ & $>7.9 \times 10^{49}$\\
\hline
\end{tabular}
\label{tab:param}
\end{table}

From the modelling, two different interpretations of the multi--wavelength light curve of GRB 140515A are possible: a pure synchrotron emission by electrons with a very flat spectrum (with $p<2$) or a multi--component model (with $p>2$). We considered only the radio, $J$, $H$ and the X-ray band emissions, since, as shown in section 3.3, the optical SDSS-$z'$ emission is strongly affected by absorption and the resulting flux is very likely underestimated.

\subsubsection{A very flat injected spectrum}

We model the observations in the $J$, $H$, radio, and X-ray bands as synchrotron radiation produced in the forward shock. In order to obtain a successful modeling we need to assume that the electron injection spectrum is a power-law with a very flat index ($p=1.67$) that extends up to some maximum Lorentz factor $\gamma_{max}$. 
Even in this case, the model cannot find a solution for the early time (very steep) X-ray emission. As suggested by e.g. Ghisellini et al. (2009) we assume that the X-ray light curve is composed by a late prompt component  decaying with time as a power--law of slope $\ge -3$ (dominant for $t_{\rm obs} \lesssim 600$~s) and by a second component interpreted as the actual X-ray afterglow emission ($t_{\rm obs} \gtrsim 600$~s). 

A good description of the detections in $J$, $H$, radio, and X-ray bands (Fig. \ref{afterglow}, dashed lines) is obtained using the parameters reported in Table \ref{tab:param}. The bump observed in the X-ray light curve is due to the onset of the afterglow, corresponding to $\Gamma_0 = 195$.

The last detection in the X-ray band shows a sudden drop of the X-ray flux. The temporal index after the break is very steep ($\alpha_{\rm late}=3.9\pm 0.6$, see table \ref{tab1}), and is not consistent with theoretical predictions from jetted outflows. As an alternative we then suggest that this steepening is due to the passage of the maximum synchrotron frequency (associated to the maximum Lorentz factor $\gamma_{max}$ of the electrons) in the X-ray band. To model the maximum frequency we assume that the ratio between the maximum and minimum Lorentz factors of the electrons is constant in time, and is equal to $\gamma_{max}/\gamma_{min}\simeq2800$. This interpretation allows us to fully describe the observations with no need for a jet break. Since the light curve does not show any break until $10^5\,$s, this time represents a lower limit on the jet break time. The corresponding lower limit on the jet opening angle and on the collimation corrected energy $E_\gamma$ are reported in Table \ref{tab:param}. This lower limit on $E_\gamma$ is consisted with the $E_{pk}-E_\gamma$ correlation \cite{EpkEg}. Consistency with this correlation, in fact, requires a jet break at times larger than $3.5\times10^5\,$s, strengthening the hypothesis that the steep break in the late X-ray light curve is not due to the jetted geometry of the outflow.

\subsubsection{Multi-component model}

An alternative interpretation could be obtained with a steeper electron injected spectrum. However, in this case it is not possible to obtain a pure synchrotron solution that can explain correctly the $J$, $H$, the X-ray emission and the radio detection. In fact, if we describe simultaneously the $J$, $H$ and radio emissions, we underestimate the X-ray light-curve. As shown also in previous sections, the peculiar shape of the X-ray light curve, characterised by an important time variability up to $t_{\rm obs} \sim 10^4$~s, suggests that the X-ray emission at those times could be probably caused by the composition of the standard afterglow emission in a forward shock scenario and some additional emission (for instance flares, or a long lasting prompt emission).

Therefore, the time of the X-ray peak must be similar to the deceleration time computed by the model. Moreover, the predicted X-ray emission cannot overcome the observed flux, since it must be a composition of the afterglow and additional emissions. In this scenario, we expect that the rise of the X-ray flux (at $t_{\rm obs}\sim 10^3$~s) corresponds to the rise of the X-ray afterglow. Adding few constrains we obtain a compatible multi-wavelength prediction of the afterglow light curve (see solid lines in Fig. \ref{afterglow}). The set of used parameters for this scenario is reported Table \ref{tab:param}.

In this scenario the X-ray afterglow prediction is below the observed data at early times, while the last observed detection becomes compatible with the expected afterglow emission. This is consistent with the hypothesis of a multi-component X-ray emission, and we can assume that only for $t_{\rm obs} \gtrsim 2 \times 10^5$~s we are observing a pure X-ray afterglow emission, not contaminated by long-lasting central engine activity. 

At the time of the Chandra observation, the X-ray afterglow flux predicted by this modelling is marginally consistent with the Chandra upper limit.
A jet break around this time would make the energetic consistent with the prediction of the $E_{pk}-E_\gamma$ correlation.
However, since a break is not strictly required by our modelling, we conservatively set a lower limit on the jet break at $t_{\rm jet}>2\times10^5\,$s. 
The corresponding lower limits on the jet opening angle and on the collimation corrected energy are listed in table~\ref{tab:param}.

\subsection{Pop III or enriched Pop II progenitor}

GRB\,140515A shows evidence of long lasting central-engine activity up to $\sim 10^4$~s after the burst event. Its redshift ($z>6$) could suggest a Pop III star progenitor. These type of massive stars ($M \ge 100 M_{\odot}$), that formed in the early universe at low metallicity ($Z \le 10^{-4}$), have been also proposed as progenitor of the so-called ultra-long GRBs, i.e.  GRB\,111209A \cite{gendre}, GRB\,121027A \cite{hou}, and GRB\,130925A \cite{evans2}.

 
 In this scenario, the long duration is the results of the time needed for the accretion and collapse mechanisms. In the hypothesis of such a GRB progenitor one should expect to detect a very low density environment with a density profile dominated by the IGM. Another expectation for such massive collapsing stars is a long-lasting blackbody emission component in their spectra, with a typical average rest-frame temperature of $kT_{\rm BB} \sim 0.5$~keV \cite{piro}. This thermal emission would be in principle detectable by BAT and/or XRT if the redshift of the event is low. 

In the case of GRB\,140515A observations do support the idea of a low density environment with negligible contribution from the host galaxy, but there are no hints for a particularly low value of the metallicity (see Section 3.4.3). Moreover, being at such a high-$z$ we do not expect to detect the blackbody component with {\it Swift} instruments. Indeed we tested this possibility (see Section 3.1) but we did not find any improvement of the fit with the inclusion of a blackbody component in the prompt emission spectrum. Therefore, the hypothesis that GRB\,140515A originated from a Pop III star (or even from a Pop II star with environment enriched by Pop III stars) is unlikely.


\begin{table} 
\small
\centering
\caption{Absorption properties of the GRBs with $z \ge 5$ (for the events marked with * the redshift was estimated photometrically). References: 1) Evans et al. 2010; 2) Perley et al. 2010; 3) Jakobsson et al. 2006;  4) Covino et al. 2013; 5) Salvaterra 2015;  6) Hartoog et al. 2014; 7) Totani et al. 2006; 8) This work.}
\begin{tabular}{cccccc}
  \hline \hline
  GRB & $z$ & log(N$_{\rm HI}$) & log(N$_{\rm H,X}$) & A$_V$ & Ref.\\
  \hline
  & & [cm$^{-2}$] & [ 10$^{21}$ cm$^{-2}$] & [mag] & \\
  \hline
  060522 & 5.11 & -- & $<160$ & -- & 1 \\
  071025 & $\le$5.2* & -- & 49 $\pm$ 19 & $< 0.54$ & 1, 2 \\
  140304A & 5.283 & -- & $<120$ & -- & 1 \\
  050814 & 5.3 & -- & $< 16.8$ & $< 0.9$ & 3, 1 \\
  131227A & 5.3 & -- & 520$^{+220}_{-190}$ & -- & 1 \\
  060927 & 5.467 & -- & $<36$ & $< 0.17$ & 4, 1 \\
  130606A & 5.913 & 19.93 & $< 30$ & $< 0.2$ & 5, 6 \\  
\hline
  120521C & 6.0$^{*}$ & -- & $< 60$ & $< 0.3$ & 5\\
  050904 & 6.295 & 21.6 & 63$^{+34}_{-29}$ & 0.15 $\pm$ 0.07  & 5,7 \\
  140515A & 6.327 & $<18.5$ & 13.5$^{+12.2}_{-10.8}$ & 0.11 $\pm$ 0.02  & 8\\
  080913 & 6.695 & 19.84 & 95$^{+89}_{-77}$ & 0.12 $\pm$ 0.03  & 5\\
  090423 & 8.26 & -- & 102$^{+49}_{-54}$ & $< 0.1$ & 5\\
  120923A & 8.5$^{*}$ & -- & $< 720$ & -- & 5\\
  090429B & 9.4$^{*}$ & -- & 140 $\pm$ 10 & 0.10 $\pm$ 0.02 & 5 \\
\hline
\end{tabular}
\label{tab:param2}
\end{table}

 \begin{figure}
\centering
\includegraphics[height=8.5cm,width=9.0cm,angle=90]{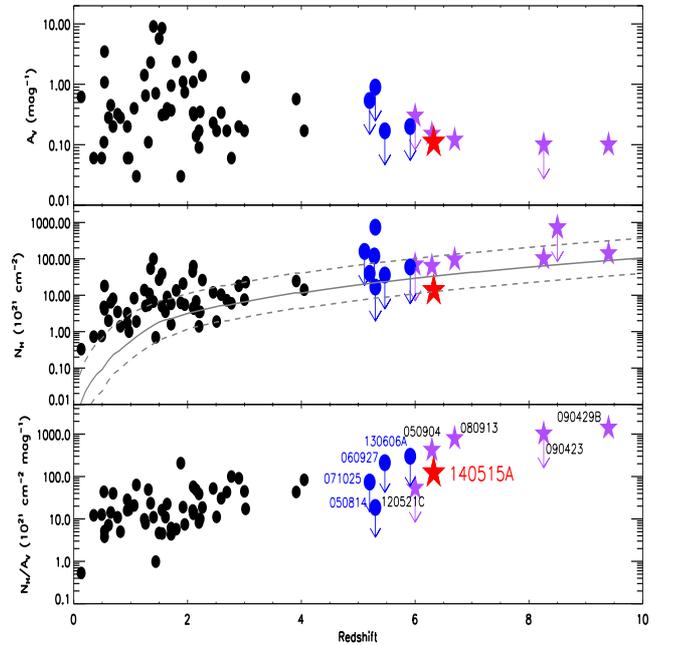}
\caption{A$_V$, N$_{\rm H}$, and N$_{\rm H}$/A$_V$ ratio as a function of redshift. Black points are from Covino et al. (2013) for events with $z \lesssim 4$, while the remaining events (blue circles, purple stars) are listed in Table \ref{tab:param2}. GRB\,140515A is marked with a red star. The solid/dashed gray lines in the middle panel represent the effect of the intervening material along the line of sight \cite[see][]{sergio15,sal15}.}
\label{znhav}
\end{figure}

\subsection{Reionization and escape fraction of ionizing radiation}

The distribution of intrinsic column densities of GRB hosts can be used to constrain the average escape fraction of ionizing radiation from the hosts \cite{chen}, under the assumption that GRB sightlines, taken as an ensemble, sample random lines-of-sight from star forming regions in GRB hosts. At intermediate redshifts ($z>2$) the sample of GRB hosts from Chen et al. (2007) indicates that only in about 5\% of all cases one expects a GRB sightline with $\log(N_{\rm HI})<18.5$. With GRB\,140515A being only 1 out of 7 GRBS with $z>6$ (and only 1 out of 3 with measured HI column densities), it appears that high redshift GRB hosts may have, on average, lower HI column densities and, hence, higher escape fractions than their lower redshift counterparts.

More quantitatively, the Kolmogorov-Smirnov test for the two distributions of HI column densities - first from Chen et al. (2007) and the second of four $z>5.9$ GRBs with measured $N_{\rm HI}$ values - shows that the two distributions are consistent with only 9\% probability. That probability raises to 30\% if GRB\,140515A is excluded. The importance of constraining the escape fractions in reionization sources is obvious, so a larger sample of $z>6$ GRBs with measured HI column densities would be highly desirable.

Such a sample would also serve as a direct test of reionization at $z>6$, where constraints from high redshift quasars become scarce. A significant advantage of GRBs over quasars is in their low or negligible bias. While bright quasars, likely, do reside in the most massive, highly biased dark matter halos, GRBs hosts at high-$z$ seem to sample the general galaxy population. Hence, constraints for the neutral hydrogen fraction obtained from the analysis of the IGM damping wing profile in the absorption spectra of GRB hosts can be expected to be more reliable than the analogous constraints from the quasar proximity zones.

In addition, constraints on the mean neutral fraction from observations of QSO proximity zones are, typically, lower limits (neutral fraction can be larger if a quasar lifetime is longer) \cite{bolton,rob13,rob15}, while constraints from GRBs are upper limits. Hence, the two observational probes are highly complementary to each other (this is demonstrated by red and orange diamonds in Fig. \ref{fig:xhi}).

\subsection{High-$z$ GRBs absorption properties}

In Table \ref{tab:param2} we report the absorption properties - neutral hydrogen column density in the host galaxy (N$_{\rm HI}$), X-ray equivalent column density (N$_{\rm H,X}$), and optical dust extinction (A$_V$) - for all known GRBs with redshift $z \geq 5$. As show in Table \ref{tab:param2}, host galaxies of high-$z$ gamma-ray bursts have small but not zero extinction, and the value of their optical extinction remains in fact rather constant (0.1 $\lesssim$ A$_{\rm V}$ $\lesssim$ 0.2 mag) over a broad range of redshifts. 

It should be noted that no large values for A$_V$ have been ever detected for GRBs at redshift $\geq 5$, and this is probably due to an observational bias, since it would be difficult to carry out optical follow-ups. Nevertheless, due to the general low metals abundances of the young galaxies at such a high-$z$, it is also plausible that large extinction are intrinsically less probable than at lower redshifts. 

While the A$_{\rm V}$ does not seem to evolve with redshift, there are no detected events with low N$_{\rm H,X}$ at  $z \ge 6$. This effect can be naturally explained by the increase of absorption of the intervening systems along the line-of-sight \cite{campana,covino,sergio15,sal15}. This mimics the evolution of the N$_{\rm H}$/A$_{\rm V}$ ratio with redshift observed in Fig. \ref{znhav} (bottom panel) up to $z \sim 10$.

\section{Conclusion}
\label{sec:concl}

We presented the multi-band spectroscopic and temporal analysis of the high-$z$ GRB\,140515A. The overall observed temporal properties of this burst, including the broad X-ray bump detected at late times, could be explained in the context of a standard afterglow model, although this requires an unusually flat index of the electron energy spectrum ($p=1.67$). Another possible interpretation is to assume that an additional component (e.g. related to long-lasting central engine activity) is dominating the X-ray emission. In the latter case, the broad band observations can be explained using a more typical value of the spectral index for the injected electron spectrum ($p=2.1$). Our modelling in this case shows that the central engine activity should cease at late times ($\sim2\times10^5\,$s), when the X-ray afterglow starts to dominate the emission. In both scenarios the cooling frequency is expected to be between the optical and the X-ray energy bands ($\nu_{c} \sim 2 \times 10^{16}$~Hz) and the average rest-frame circum-burst extinction (A$_{\rm V} \sim 0.1$) resulted to be typical of high-$z$ bursts. 

Our detailed spectral analysis provided a best estimate of the neutral hydrogen fraction of the IGM towards the burst of $x_{HI} \leq 0.002$ and a conservative upper limit of the HI abundance in the GRB host galaxy of $N_{\rm HI} \lesssim 10^{18.5}$ cm$^{-2}$. These values are slightly different from the ones estimated by Chornock et al. (2014b). In addition, the spectral absorption lines observed in our spectra are the weakest lines ever observed in GRB afterglows \cite{ugarte0}, suggesting that GRB\,140515A happened in a very low density environment. However, our upper limits on the gas-phase abundances, coupled with the fact that we cannot establish the exact metal-to-dust ratio, do not allow us to distinguish between metallicity in the range of $10^{-4} < [Z/H] < 0.1$. This makes the possible Pop III star origin for GRB\,140515A uncertain and doubtful.

For all high-$z$ GRBs the contribution of the host galaxy was not negligible (Table \ref{tab:param2}). GRB\,140515A is the first case when this does not happen, allowing us to give the the best observational constrains on a theoretical model at $z>6$. 

\begin{acknowledgements}
This research has been supported by ASI grant INAF I/004/11/1. This work made use of data supplied by the UK Swift Science Data Centre at the University of Leicester.
\end{acknowledgements}



\newpage


\end{document}